# DESIGN AND VALIDATION OF SAFETY CRUISE CONTROL SYSTEM FOR AUTOMOBILES

Jagannath Aghav and Ashwin Tumma

Department of Computer Engineering and Information Technology,
College of Engineering Pune,
Shivajinagar, Pune, India
`{jva.comp, tummaak08.comp}@coep.ac.in`

## ABSTRACT

*In light of the recent humongous growth of the human population worldwide, there has also been a voluminous and uncontrolled growth of vehicles, which has consequently increased the number of road accidents to a large extent. In lieu of a solution to the above mentioned issue, our system is an attempt to mitigate the same using synchronous programming language. The aim is to develop a safety crash warning system that will address the rear end crashes and also take over the controlling of the vehicle when the threat is at a very high level. Adapting according to the environmental conditions is also a prominent feature of the system. Safety System provides warnings to drivers to assist in avoiding rear-end crashes with other vehicles. Initially the system provides a low level alarm and as the severity of the threat increases the level of warnings or alerts also rises. At the highest level of threat, the system enters in a Cruise Control Mode, wherein the system controls the speed of the vehicle by controlling the engine throttle and if permitted, the brake system of the vehicle. We focus on this crash area as it has a very high percentage of the crash-related fatalities. To prove the feasibility, robustness and reliability of the system, we have also proved some of the properties of the system using temporal logic along with a reference implementation in ESTEREL. To bolster the same, we have formally verified various properties of the system along with their proofs.*

## 1. INTRODUCTION

With the advent of an era of new technological advances and developments, there has been a considerable growth in almost all the facets; being it human population or the industries. In accordance with the same, there also has been an abundant and herculean increase in the number of vehicles or automobiles on the roads. Consequently, this increase of vehicles has led to a alarming growth of the fatal road accidents throughout the globe. Statistics depict that more than 2.2% of the total deaths recently have occurred because of the road crashes which could have been prevented. Also, if the same statistics are at play in future, then the World Health Organization by 2020, road fatalities will be the third highest threat to the public health, outranking most of the dangerous health problems [20].

The above discussion, clearly brings into light that, today, the need of the hour is to curb the rate of road fatalities. In light of the same, as a solution to the stated issue, we have proposed a safety cruise control system which addresses the problem of minimizing the number of vehicle crashes due to erroneous controlling of the vehicle, and thereby decreasing the road accidents. Safety Cruise Control System for Automobiles with ESTEREL Implementation and Validation is our proposal for safety system for automobiles wherein, the automobile will be equipped with a Safety System, which will alert the drivers when there is a potential for crash. It consists mainly of a safety algorithm and a Cruise Control System. The goal is to reduce the number and severity of automobile fatalities and crashes. The system is broadly classified in two sub-systems:

- Safety System
- Cruise Control System

These form the two major working units of the system. The architecture of the system is shown in Figure 1.

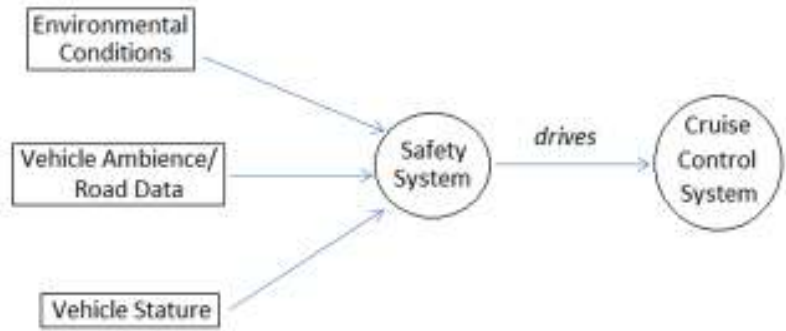

Figure 1. Subsystems of the Architecture

Figure 1 illustrates the architecture of the system in brief. Initially, the safety system considers the environmental conditions in which the vehicle is operating, plus it collects data from the ambience of the vehicle, and then checks the current stature of the host vehicle. It then analyses the acquired data and then reports to the driver accordingly. The reports to the driver are sent through the Driver Vehicle Interface. The driver can also interact with the system via this interface. Later, if the safety system discovers a potential of a crash, it then drives the Cruise Control system by asking it to come into play and control the operations of the vehicle. In this way, since the Cruise control system will have the control of the vehicle in crash-probable circumstances, the chances of safeness rise as the crash will be mitigated in the cases, where it is possible to shun the crash. The details of the working of each subsystem are presented in the subsequent sections.

In this paper, we propose the safety system along with its implementation in a synchronous programming language named ESTEREL. We also prove the robustness and reliability of the system by stating and proving certain properties of the system. Initially, we will state some of the specifications of the system by making use of temporal logic, and will then justify by formal verification that our implementation conforms to the specification stated; thereby warranting the pragmatic genre of the system.

Rest of the paper is organized in the following manner. Section 2.1 discusses the Safety System and its intricacies. Section 2.2 introduces the Cruise Control system. Section 2.3 explicates the details of the architecture of the system. Section 2.4 provides a snippet of reference implementation of the safety system in ESTEREL. Section 2.5 presents the specification in temporal logic along with the formal verification of the ESTEREL modules. Section 3 presents the conclusions of the paper.

## 2. THE SYSTEM DESIGN

This section explicates the details of the system, with throwing special light on drafting the specifications and then verifying them for the proposed system.

### 2.1. Safety System

The Safety System forms the heart of the Safety Cruise Control System [19], [13], [12]. It consists of a sensor (Section 2.1.1) that gathers data from the vehicle's ambience. At each instance of time, here each instance of time can be mapped to each clock tick, the sensor gets the new roadway data and this data is then analysed by the safety algorithm to check it against the predefined safety parameters. The concept of pre-defined parameters will be explained in next section. If the current host vehicle conditions are such that they are in close physical proximity to the threshold limit of the safety parameters, then the system sends an alert to the driver that there is a potential for a crash with the lead vehicle or an arbitrary object. Also, if the current circumstances are such that there is a high probability of crash or any other accident, the safety algorithm instructs the Cruise Control System to take over the controlling of the vehicle. Details of Cruise Control System are documented in Section 2.2.

### 2.1.1. Sensor Details

The Safety System mentioned above makes use of a sensor to collect the data of various parameters from the vehicle's environment. Our proposal includes employment of a sensor (off-the-shelf-component) named Forward Looking Automotive Radar Sensor. This sensor perfectly suffices our purpose since it is specially designed to be used in Intelligent Cruise Control Systems and Collision Warning Systems. Following paragraph talks about the specifications of the sensor.

*A Forward Looking Automotive Radar Sensor:* This sensor available from [10] is a specially built sensor for intelligent cruise control and forward looking collision warning systems. They are used to collect information about traffic and obstacles in the roadway ahead. Few of the distinguishing features of this sensor are:

- It correctly identifies a lead-vehicle being followed, constantly distinguishing between lead vehicle and competing vehicles and roadside objects.
- Report the distance and relative speed of the lead vehicle to platform vehicle speed control unit.

The specifications of the sensor are given in Table 1.

Table 1. Sensor Performance Specifications

| Characteristic | Value |
| --- | --- |
| Operating Frequency | 76-77 GHz (MMW) |
| Range | 3-10+ meters |
| Range Accuracy | << 0.5 meters |
| Relative Speed | +/- 160 Km/h |
| Field of View (Azimuth) | 9 Degrees |
| Interface | SAE J1850, RS-232, High Speed Parallel |

The sensor specifically makes use of algorithms to interpret the transmitted and received radar signals to determine the distance, relative speed and azimuth angle between host vehicle and the vehicle or object ahead of it in the lane. The ESTEREL Module gets this data through interfaces and then applies its algorithm on it.

## 2.1.2. Safety Algorithm

The sensor collects the data from the environmental conditions and current ambience of the vehicle at each instance of time and forwards it for analysis to the ESTEREL Module. The Safety Algorithm then compares the values of the various parameters in the received data with the set of predefined parameters. If the received values are close to the threshold limit of that particular parameter, then the algorithm emits a signal to the driver through the Driver Vehicle Interface, that there is a potential for a crash with the lead vehicle or an object in front of the host vehicle in the lane. We first discuss the parameters that are taken into consideration to identify the potential threat, the different proposed choices to set the predefined parameters and how the data from the sensor is analysed.

*Predefined Parameters:* Physical parameters of the vehicle, roadway and other objects are taken into account which assists us in identification of potential for a crash or any other threat. The parameters are: distance, relative speed and azimuth. Distance is the distance between the host vehicle and the lead vehicle or an object in the lane. Relative Speed is the speed of the lead vehicle with respect to the host vehicle. Azimuth field-of-view of the camera is the span of the angle between two boarders that falls in the sensitivity of the installed sensor device. All these parameters are illustrated in the Figure 2.

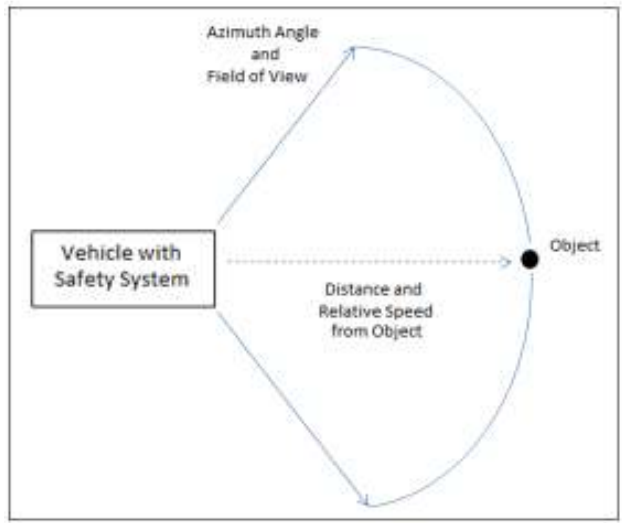

Figure 2. Parameters under consideration for front crashes

The threshold values of these parameters are set in the safety algorithm. We term these parameters as Predefined Parameters. They are altered as per the choices mentioned in next paragraph.

*Choices to set the Predefined Parameters:* We propose three different choices to the user/driver for setting the values of the predefined parameters mentioned in the previous paragraph, viz.

1) The manufacturer can fabricate the default values of these parameters.
2) Driver can customize these values as per his driving habits.
3) The system can automatically learn its environmental conditions and set the parameters accordingly.

Initially, when the system is being configured the manufacturer can set the threshold values of the parameters. These values could be set with certain generic conditions in mind. For example, the overall condition of the roads in the country, overall traffic statistics, etc. The user has an option of retaining these values or customizing them according to his driving habits and convenience. Also, these values can be adapted according to the environmental conditions in which the vehicle is running. For instance, if it is raining, the values can be adjusted accordingly so that the system gives an alert at a considerable safe distance from the remote object, or if the vehicle is travelling on a road which has dense fog, the values need to be altered in such a way that they suit the current environment of the vehicle and the alert is at such a distance that the vehicle can be controlled safely to shun the crash. For monitoring the environmental conditions, micro-condensed sensors or sensors that can judge their ambience can be used. We assume that this data is also sent to our ESTEREL Module which takes into account the climatic conditions of the vehicle.

*Data from the Sensor:* The sensor collects real-time parameters from the vehicle's ambience. The parameters that are taken into consideration are: distance, relative speed and azimuth. The definitions of these parameters are the same as explained above. These parameters are sent to the safety algorithm by means of an interface, and then the algorithm uses them for further analysis.

### 2.2. Cruise Control System

Cruise Control System [15], [18] forms the second subsystem of Safety Cruise Control for Automobiles. As mentioned in the previous section, the safety algorithm sends appropriate notifications to the driver whenever there is a potential for a crash or any other threat. However, if the vehicle is in close physical proximity to the lead vehicle or any object, it might be the case that the driver reaction time is not so fast that it can preempt the crash. In such cases, the algorithm sends a signal of a high level threat and instructs the Cruise Control Subsystem to take over the control of the vehicle and reduce the speed by controlling the engine throttle and if permissible the brake system of the vehicle, thereby making the best possible attempt to pre-empt the crash and avoid any further injuries or fatalities. An appropriate alert is sent to the driver through the Driver Vehicle Interface; so that the driver can also keep a track of the fact, that the Cruise control system in his vehicle is playing the role of the avoiding the crash.

### 2.3. Architecture

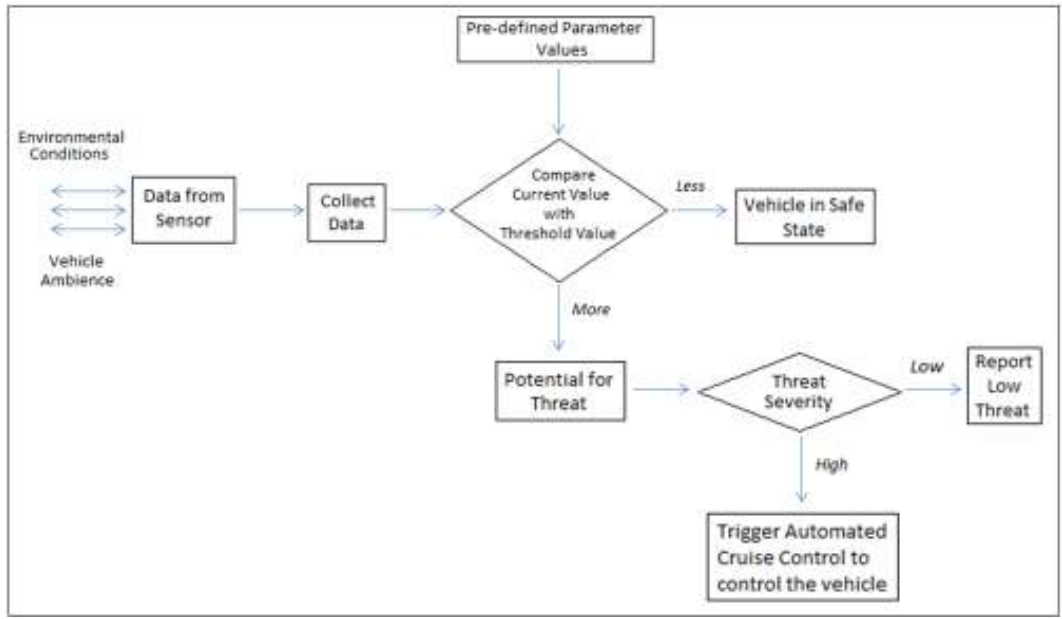

Figure 3. Detailed Architecture of the System

The previous sections described about the two subsystems of ESTEREL Implementation and Validation of Cruise Controller. This section highlights the architecture of the entire system. As shown in the Figure 2, the input to the system will be from the Automotive Radar Sensor. The sensor makes use of algorithms to interpret the transmitted and received radar signals to determine the distance, relative speed and azimuth between host vehicle and the vehicle or object ahead of it in the lane. These parameters are used as an input to the safety algorithm. The algorithm in turn compares these currently available values with a set of threshold values to check if there is a potential for a crash and notifies the driver accordingly to take appropriate actions. If the level of threat is high, then the system enters in Cruise Control Mode and takes over the control of the system to mitigate the threat.

Figure 3 shows a detailed architecture of the system. The current environmental conditions and the data from the vehicle's ambience like relative distance, speed and azimuth angle are taken as input to the sensor. The sensor, then passes the current values to the safety system. These current values are compared with the set of the pre-defined values of those respective parameters. The predefined parameters can be overwritten by the values from the Environment Monitor. If the current values of the parameters are less than the threshold limit, then the vehicle is in safe state and can proceed safely. But the moment the values surpass the threshold limit, an alert is raised. Now, depending on the threat severity, the system chooses its mode of operation. If the level of threat is low, then an appropriate notification or alert is sent to the driver and if the level is high, the system enters the Cruise Control Mode to take over the controlling of the vehicle.

*Driver Vehicle Interface:* The Driver Vehicle Interface (DVI) will be the means by which the driver can get the visual warnings. We define different levels of alarms based upon the severity of the threat. As the sensor can identify objects with their distance, if the object is far enough, then a minimal warning can be issued, and as the distance between the host vehicle and lead vehicle reduces, the level of warnings can be intensified. Audio warnings depending on the threat severity can be employed, i.e. different tones of varying length for different threat levels. These tones can also be customizable by the driver. This interface will also provide means from which the driver can access the feedback given to him by the system like the notifications and alerts sent while his is driving the vehicle.

### 2.4. ESTEREL Module

### 2.4.1. ESTEREL

Many real time applications demand reactive systems. Reactive systems are the ones which continuously react to their input signals and optionally produce output signals which are used by other systems. Such systems need to have support for Control Handling. Control Handling deals with producing discrete output signals for the input signals. ESTEREL [17], [3], [8], [4] is a synchronous and imperative concurrent programming language that is used for programming reactive systems. It also provides a compiler that translates ESTEREL programs into their associated finite-state machines. It provides support for sensors and signals that can be easily received and emitted by the modules. It is easily employed in applications, which need communication of data in its subsystems. This communication is achieved by means of broadcasting of signals. Various modules of a system use these broadcasted signals for sharing the data available with them.

We chose ESTEREL for our implementation purposes because; our system needs synchronous communication of various parameters from the vehicle's ambience (collected by the sensor) and the predefined parameters to the safety algorithm. This can be easily achieved by building different modules for different subsystems which work in perfect synchrony. The subsystems can send signals to each other and can work in parallel. Also, the alerts need to be sent to the driver through the Driver Vehicle Interface. These alerts can be sent by emitting signals from the appropriate modules. ESTEREL being an imperative concurrent programming language allows synchronous communication between the modules, which perfectly models the real time behaviour of the system and also simplifies the embedding of these modules in related hardware circuitry. Section II-D2 presents snippets of the various modules for building our system. A formal verification of some of the modules is also presented in Section II-E.

### 2.4.2. ESTEREL Implementation of Proposed System

The language constructs and coding conventions can be found in [1], [2]. Following are the snippets of ESTEREL Codes for implementation of our system. In this reference implementation we do not consider the parameter named azimuth field of view. Separate modules are built for the different subsystems shown in Figure 1. These modules communicate with each other by broadcasting the signals.

```
module SAFETY_SYSTEM :
Run SET_PREDEFINED_VALUES;
Run ROAD_DATA;
||
Run HOST_VEHICLE;
||
Run DRIVER_ALARM;
end module
```

This module forms the main algorithm which invokes other modules and runs them in parallel. In the first step it runs the SET_PREDEFINED_VALUES module which is used to set the predefined values of the various parameters. Further, it invokes the modules ROAD_DATA, HOST_VEHICLE and DRIVER_ALARM and runs them in parallel throughout the life of the program.

```
module SET_PREDEFINED_VALUES :
var distance := 12 : integer;
% While Fabrication
var speed := 20 : integer;
output PreDefinedDistance : integer;
output PreDefinedSpeed : integer;
Run ENVIRONMENT_CHECK;
||
present
case rain do
distance := 10;
speed := 18;
case mist do
distance := 8;
speed := 17;
case normal do
distance := 5;
speed := 20;
Run DRIVERINPUT ;
||
await InputDistance;
||
await InputSpeed;
if distance < ?InputDistance then
distance := ?InputDistance;
emit PreDefinedDistance(distance);
```

```
end if
if speed < ?InputSpeed then
speed := ?InputSpeed;
emit PreDefinedSpeed(speed);
end if
end module
```

This module, called by SAFETY_SYSTEM, sets the values of the predefined parameters (The values of predefined parameters used in these modules are specimens only. They may differ in practical implementations). The manufacturer can set the values whilst fabrication or these values can be user-driven and environment adaptive. The module, first calls module ENVIRONMENT_CHECK which returns the climatic condition of the vehicle's ambience. Based on these conditions the values of parameters are set. These values are then compared with those obtained from the driver. Finally the safest values are assigned to the parameters.

```
module ENVIRONMENT_CHECK :
sensor climate;
% Get Data from the sensor and
% output the climatic condition
end module
```

The environmental conditions in which the vehicle is running are analysed by this module. We assume that the sensors used for building this module are capable of sensing the ambient conditions. It returns the appropriate climatic condition to module SET_PREDEFINED_VALUES.

```
module DRIVERINPUT :
output InputDistance : integer;
output InputSpeed : integer;
var distance := ?enteredDistance;
emit InputDistance(distance);
var speed := ?enteredSpeed;
emit InputSpeed(speed);
end module
```

This module accepts the values of parameters from the driver through the Driver Vehicle Interface and passes these values to the SET_PREDEFINED_VALUES module. This completes the steps required for initiating the system. Now, the system starts running by executing ROAD_DATA, HOST_VEHICLE and DRIVER_ALARM in parallel. These continue to run till the vehicle's engine is turned off.

```
module ROADDATA:
input distance, speed, SAMPLE_FREQ,
STOP_VEHICLE,
RUNNING;
output DistanceSignal, SpeedSignal;
weak abort
every immediate SAMPLE_FREQ do
present RUNNING then
loop
present [distance and speed] then
emit DistanceSignal
||
emit SpeedSignal;
end present;
pause;
end loop;
```

```
end present;
end every;
when STOP_VEHICLE;
end module
```

ROAD_DATA module fetches the data from the sensor and makes it available for other modules. The samples of the data are collected with the sampling rate determined by SAMPLE_FREQ. As the number of revolutions of the wheel increase, the sampling rate increases accordingly.

```
module DRIVER_ALARM:
output Alert;
var criticalDistance := 4 : integer;
var criticalSignal := 10 : integer;
await DistanceSignal;
||
await PreDefinedDistance;
||
await SpeedSignal;
||
await PreDefinedSpeed;
if ?DistanceSignal <= ?PreDefinedDistance or
?SpeedSignal <= ?PreDefinedSpeed then
if ?DistanceSignal <= criticalDistance
or
?SpeedSignal <= criticalSpeed then
% Enter Cruise Control Mode;
emit Alert(1);
else
% Raise Lower Level Alarm
emit Alert(0);
end if
end if
end module
```

DRIVER_ALARM module sends an alert to the driver about the threats. It gets the roadway data and the values of the predefined parameters. If the values of parameters from the roadway data drop below a threshold value(criticalDistace and criticalSpeed), then a lower level alert is sent to the driver and if the values drop below a certain critical value then the module sends a higher level alert to the driver. Alert(1) signifies that there is a high level threat, Alert(0) signifies a lower level threat.

```
module HOST_VEHICLE:
input LowAlert, CruiseControlAlert;
output LowNotification, CruiseControlMode;
present
case LowAlert do
emit LowNotification;
case CruiseControlAlert do
run Cruise;
end present;
end module

module CRUISE:
output CruiseControlMode;
emit CruiseControlMode;
end module
```

This module sends notification to the driver through the Driver Vehicle Interface when there is a lower level threat and if a higher level threat is detected it calls the CRUISE module which emits the signal for the Cruise Control mode and takes over the control of the system.

```
module CRUISE_CONTROL:
input SAMPLE_FREQ, CruiseControlMode;
output ControlEngine, ControlBrake,
NotifyDriver;
every immediate SAMPLE_FREQ do
present CruiseControlMode then
emit ControlEngine
||
emit ControlBrake
||
emit NotifyDriver;
end present ;
end every;
end module
```

The CRUISE_CONTROL module handles the Cruise Control Mode of the system. Whenever the vehicle enters in a critical threat region, this mode is activated. It sends appropriate control signals to the hardware circuitry in the vehicle to control the engine throttle and the brake system. ControlEngine, ControlBrake and NotifyDriver signals are further used by the vehicle's control circuitry for actionable control of the vehicle.

## 2.5. Formal Verification of ESTEREL Modules

This section presents the formal verification [16], [5] of the Esterel Modules presented above. We verify by using FSM minimization and checking the status of outputs and verifying the properties [5]. We employ a formal verification of this system which allows us to test the conformance of the design with specification. For verification, we will use Xeve Verification environment from the ESTEREL toolset [7], [9] to verify the formal correctness of the Esterel modules presented in the previous sections. The method of model checking is to represent the design as a reduced finite automata by FC2TOOLS [6], [11] using the concept of bisimulation [14]. The states are shown by circles, inputs by "?" and outputs by "!". Fig. 4 shows the reduced automata for the RoadData module.

*Checking output signals and verifying properties:* Initially we specify the properties in terms and notations of temporal logic [21], [22] and then provide its explanation. Theoretical details of checking output signals and verifying their properties are available at [5]. Fig. 5 shows the snapshot of Xeve verifying the RoadData module. Consider the following property:

*p1: RunningState $\rightarrow \Diamond ValuesBroadcasted$*

*When the vehicle is in RUNNING state, then for each SAMPLE_FREQ the distance and speed are taken as inputs and their values are broadcasted to other modules.*

Notice that p1 claims that when the vehicle is on and is in running state then for each sampling frequency SAMPLE_FREQ, it shall get the current values of distance and speed and broadcast them to the other modules. As shown in Fig. 5, we can verify p1 by setting RUNNING and SAMPLE_FREQ as "always present" (marked as red in left window), while setting the STOP VEHICLE to "always absent". DistanceSignal and SpeedSignal are set to red which means they are to be checked if they are "possibly emitted". On clicking "Apply", Xeve shows that DistanceSignal and SpeedSignal are Always Emitted, and by this we can conclude that the current values of distance and speed are broadcasted. Similarly consider property

*p2: VehicleStop → □¬(EmitSignal)*

*When the vehicle stops, the system should stop emitting signals and halt.*

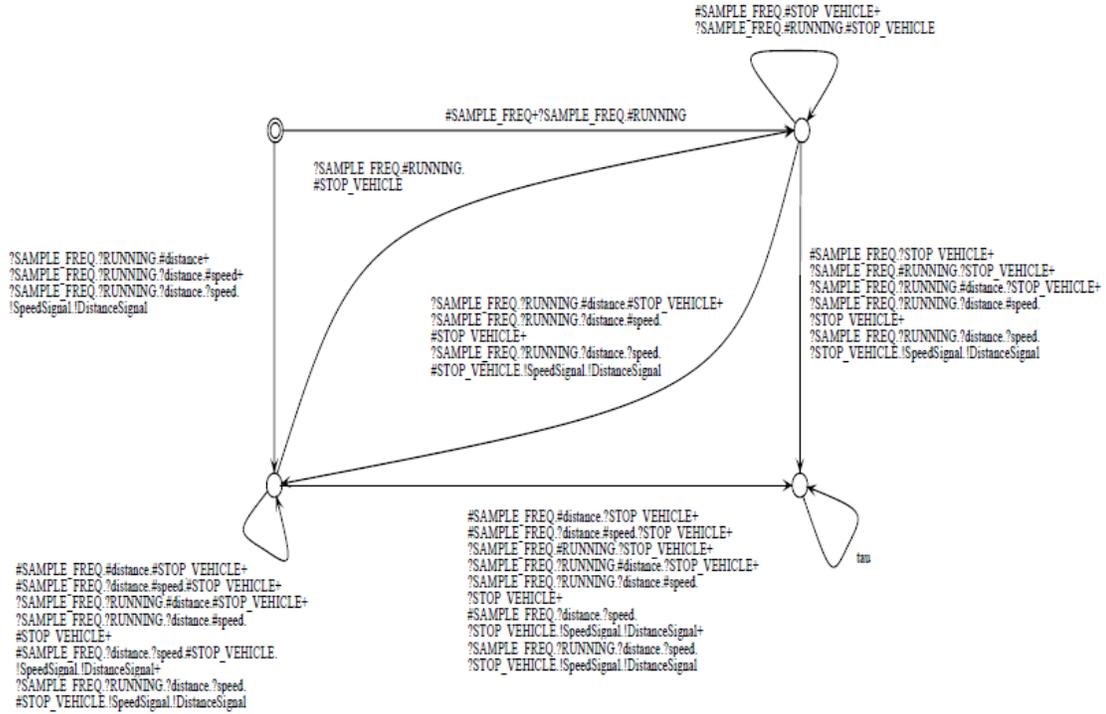

Figure 4. Reduced Finite Automata of module RoadData obtained from FC2TOOLS

Fig. 6 verifies this property, wherein, now we set STOP VEHICLE as "Always present" and RUNNING as "Always absent". If this condition occurs, then we say that the vehicle has stopped running and is halted. Consequently, (as shown in right window of Fig. 6) the DistanceSignal and SpeedSignal are also not emitted and the system halts which also verifies p2.

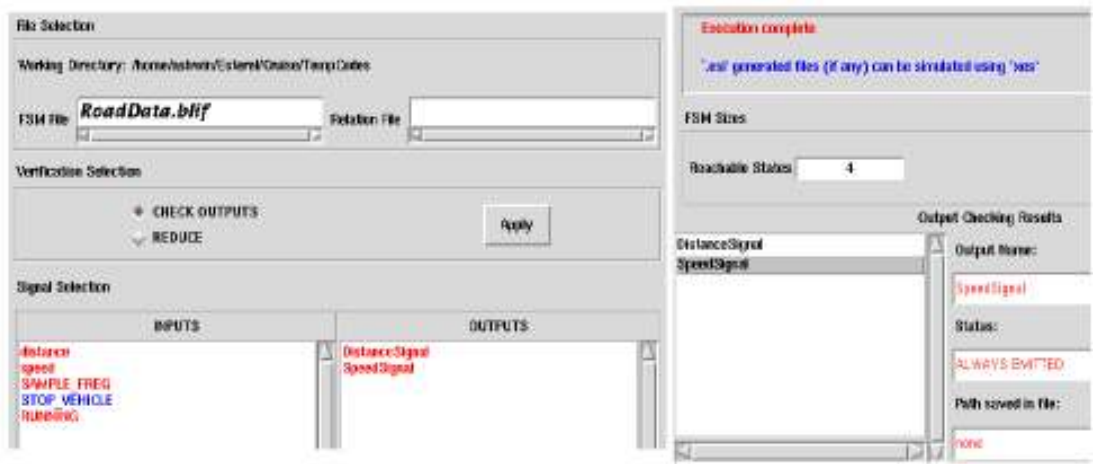

Figure 5. Snapshot of Xeve Verification for RoadData module when vehicle is running

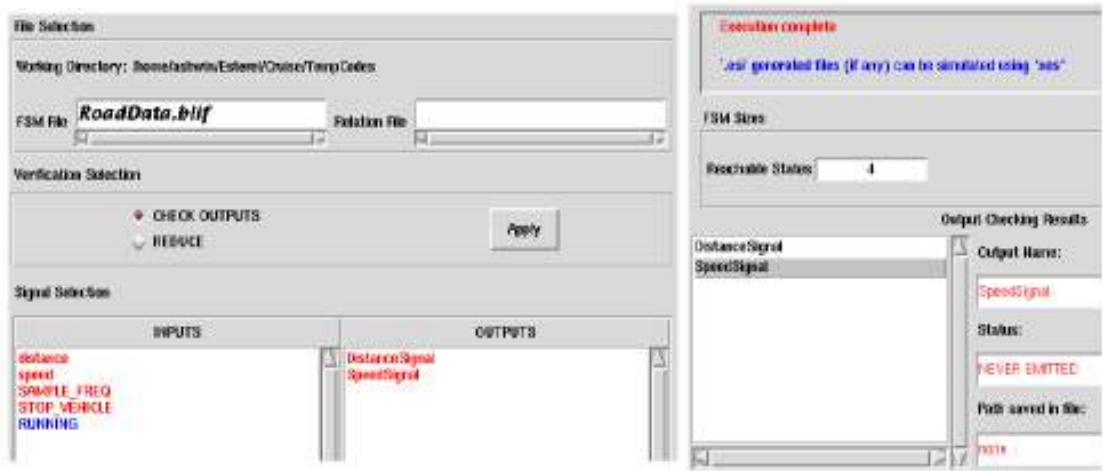

Figure 6. Snapshot of Xeve Verification for RoadData module when vehicle is Stopped

We now consider the HostVehicle module for verification. The vital property that needs to be verified in this module is that,

*p3: PotentialForCrash → □ (CruiseControlMode)*

*Whenever the vehicle goes in a close physical proximity to the lead vehicle or object, the system enters the Cruise Control Mode.*

This property claims that whenever the vehicle's speed and distance cross the threshold values of alert as set by the system, or in other words, is in close physical proximity to the lead vehicle or the object, the safety system should take over the control of the system and enter the Cruise Control Mode. Fig. 7 shows snapshot of verification of this property, in which we set the CruiseControlAlert Signal to be "Always present", which in turn says (as shown in right window of Fig. 7) that CruiseControlMode Signal is emitted and the system will enter the appropriate mode, thereby avoiding the immediate crash with the lead vehicle or object. Fig. 8 also shows that, when the system is to enter in the Cruise Control Mode, a LowLevelAlert is never emitted, which consequently verifies that the system works ideally in cases of where there is an immediate probability for crash.

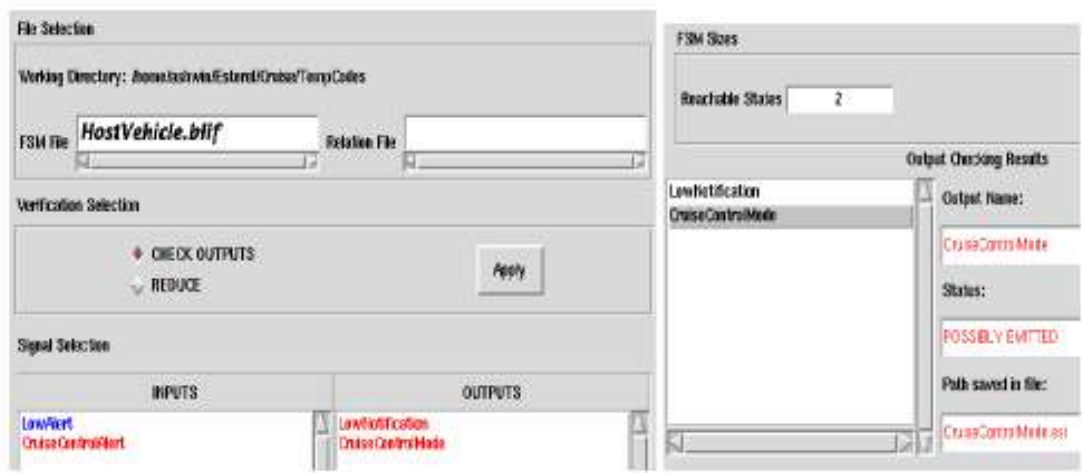

Figure 7. Snapshot of Xeve Verification for HostVehicle module with presence of CruiseControlAlert Signal

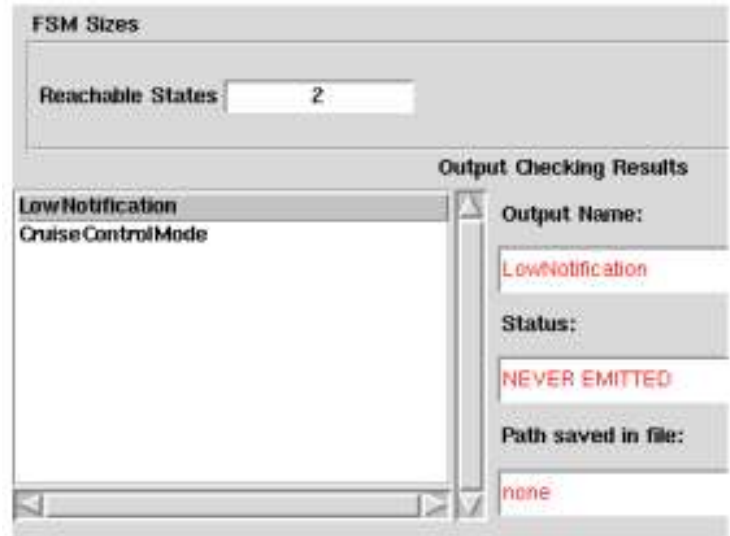

Figure 8. Snapshot of Xeve Verification for HostVehicle module showing the status of LowNotification Output Signal

*p4: CruiseControlMode → □ (ControlEngineCircuitry ∧ NotifyDriver)*

*When in Cruise Control Mode, the system takes over the control of the brake and engine circuitry and notifies the driver about the same*

The above property p4 says that, whenever the system works in Cruise Control mode, it takes over the brake circuitry of the vehicle and controls the engine throttle to lower the pace of the vehicle and mitigate the probable crash with the lead vehicle or object. Also, the driver needs to be informed about the same through the DVI panel. Fig. 9 illustrates the snapshot of CruiseControl module in Xeve verification, wherein, we have set the SAMPLE_FREQ and CruiseControlMode signals to be "Always present"and one of the output signal (NotifyDriver) to be checked if it is "possibly not emitted", while setting other output signals to be checked if they are "possibly emitted". We observe that, no matter the check of whether the signals are possibly emitted or not, our module Always Emits the ControlBrake, ControlEngine and NotifyDriver signals, from which we can conclude that the system behaves as intended in the Cruise Control Mode also.

The above formal specification supports the formal reasoning, which, because it can be checked by machine can be made very reliable indeed. Formal Verification thus guarantees the robustness and reliability of our reactive system, since it is possible to calculate the truth or falsehood of the specification by simply checking the status of output signals or traversing along the states of the finite automata, thereby proving the logical correctness of the system. Because of these distinguishing characteristics, its simulation on hardware is also simplified, which also assists in the reduction of development time and efforts.

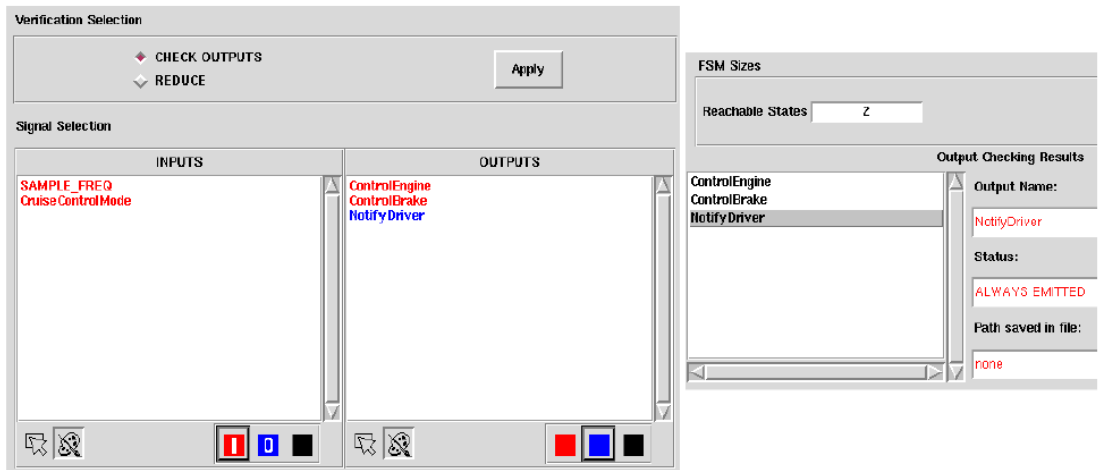

Figure 9. Snapshot of Xeve Verification for CruiseControl module showing ALWAYS EMITTED status for NotifyDriver Output Signal

*Our Platform for Verification:* For verification purposes, we have used ESTEREL and Xeve[7], [9], [6], [11] on GNU/Linux Ubuntu 10.04 Lucid Lynx running on Dell Studio 1531 with Intel(R) Core(TM)2 Duo CPU T6400 @ 2.00GHz speed having Cache size 2048 KB and two CPU cores.

## 3. CONCLUSIONS

In this paper we have proposed a reactive system based implementation of Cruise Controller using Synchronous programming language. The Safety System and the Cruise Control system help in mitigating the crashes of the host vehicle with other vehicles and objects. The system adapts according to the environmental conditions thereby increasing the safeness of the vehicle in almost all climatic conditions. The implementation responds faster as ESTEREL logically takes no time as compared to other existing systems and lucidly suits the hardware implementation. A formal verification of the system along with verification of different properties is also done to assure the correctness of the system and ensure the robustness and reliability of the system. Encapsulating, the system will assist to improve the safeness of vehicles and shall reduce the vehicle crashes on roads.

**Authors:**

**Dr. Jagannath Aghav** is Professor in the Department of Computer Engineering and Information Technology at College of Engineering (COEP) Pune, India. Contact him at: jva.comp@coep.ac.in

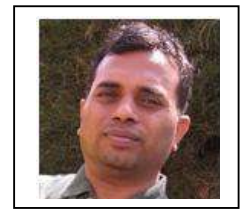

**Ashwin Tumma** is a senior undergraduate student at the Department of Computer Engineering and Information Technology at College of Engineering, Pune, India. Contact him at: tummaak08.comp@coep.ac.in

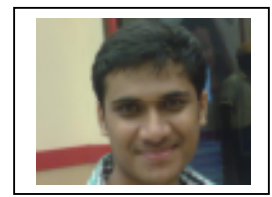